\pdfoutput=1
\documentclass[sigconf,screen=true]{aamas} 

\usepackage[T1]{fontenc}
\usepackage[utf8]{inputenc}

\AtBeginDocument{%
  \providecommand\BibTeX{{%
    \normalfont B\kern-0.5em{\scshape i\kern-0.25em b}\kern-0.8em\TeX}}}

\setcopyright{none}

\usepackage{subfig}
\usepackage{float}
\usepackage[capitalise, noabbrev]{cleveref} % Automatically include e.g. "Section" in \ref
\usepackage{soul} % strikethrough text
\usepackage{colortbl} % coloured tables
\definecolor{lightgray}{gray}{0.925}
\usepackage{multirow}
\usepackage{array}
\usepackage{tabularx}
\usepackage{longtable}
\usepackage{pdflscape}
\usepackage{pifont}
\usepackage{arydshln}
\usepackage{placeins}
\usepackage{xfakebold}
\usepackage {xcolor}
\usepackage{enumitem} % Better control of lists and their spacing
\setlist{nolistsep} % Remove vertical spacing in lists

\setcopyright{ifaamas}
\acmConference[AAMAS '22]{Proc.\@ of the 21st International Conference
on Autonomous Agents and Multiagent Systems (AAMAS 2022)}{May 9--13, 2022}
{Online}{P.~Faliszewski, V.~Mascardi, C.~Pelachaud,
M.E.~Taylor (eds.)}
\copyrightyear{2022}
\acmYear{2022}
\acmDOI{}
\acmPrice{}
\acmISBN{}

\acmSubmissionID{151}


\begin{document}
% Title portion
%\title{Multimodal analysis of hand gesture properties predictability}
\title[Multimodal analysis of the predictability of hand-gesture properties]{Multimodal analysis of the predictability\\ of hand-gesture properties}

\author{Taras Kucherenko}
\affiliation{
\institution{KTH Royal Institute of Technology}
  \city{Stockholm}
  \state{}
  \country{Sweden}
 }
\email{taras.svitozar@gmail.com}

\author{Rajmund Nagy}
\affiliation{
\institution{KTH Royal Institute of Technology}
  \city{Stockholm}
  \state{}
  \country{Sweden}
 }
\email{rajmundn@kth.se}

\author{Michael Neff}
\affiliation{
\institution{University of California}
  \city{Davis}
  \state{}
  \country{United States}
 }
\email{mpneff@ucdavis.edu}

\author{Hedvig Kjellström}
\affiliation{
\institution{KTH Royal Institute of Technology}
  \city{Stockholm}
  \state{}
  \country{Sweden}
 }
\email{hedvig@kth.se}

\author{Gustav Eje Henter}
\affiliation{
\institution{KTH Royal Institute of Technology}
  \city{Stockholm}
  \state{}
  \country{Sweden}
 }
\email{ghe@kth.se}

% Currently planned authors:
% Taras Kucherenko, Rajmund Nagy, Michael Neff, Hedvig Kjellström, Gustav Eje Henter.

%\renewcommand\shortauthors{Zhou, G. et al}

\begin{abstract}
Embodied conversational agents benefit from being able to accompany their speech with gestures. Although many data-driven approaches to gesture generation have been proposed in recent years, it is still unclear whether such systems can consistently generate gestures that convey meaning. We investigate which gesture properties (phase, category, and semantics) can be predicted from speech text and/or audio using contemporary deep learning. In extensive experiments, we show that gesture properties related to gesture meaning (semantics and category) are predictable from text features (time-aligned FastText embeddings) alone, but not from prosodic audio features, while rhythm-related gesture properties (phase) on the other hand can be predicted from audio features better than from text. These results are encouraging as they indicate that it is possible to equip an embodied agent with content-wise meaningful co-speech gestures using a machine-learning model.
\end{abstract}

\newcommand{\rn}[1]{\textcolor{red}{#1}}

\newcommand{\rnst}[1]{\textcolor{red}{\st{#1}}}



\pagestyle{fancy}
\fancyhead{}

\maketitle

\section{Introduction}

\begin{figure}
\centering
  \includegraphics[width=0.5\textwidth]{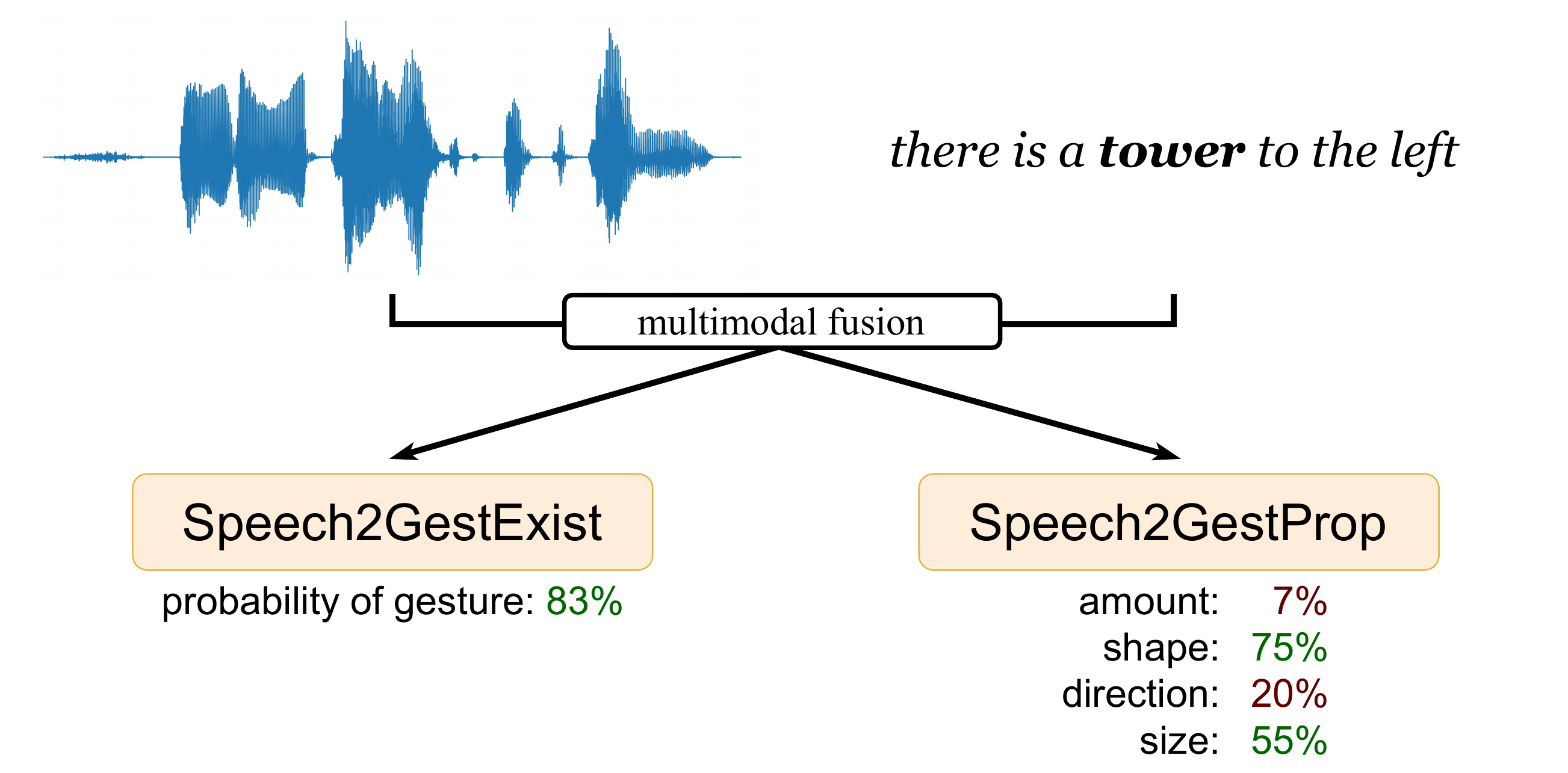}
  \caption{An illustration of the problem we study.}% framework.}
  \Description{A picture of an audio waveform and the example sentence "there is a tower to the left" (with tower in bold) are connected to a labelled box that says "multimodal fusion". Arrows point from this box to two other boxes: one titled "Speech2GestExist", with "probability of gesture: 83 percent" written under it, and another titled "Speech2GestProp", with the following text under it: "amount: 7 percent, shape: 75 percent, direction: 20 percent, size: 65 percent".
  }
  \label{fig:teaser}
  \vspace{-3mm}
\end{figure}

Verbal and nonverbal communication are important and complementary components of embodied human communication. In human communication, speech is typically accompanied by \emph{co-speech gestures} or \emph{gesticulation}, performed by the %motion of the 
hands, head, and occasionally the body.
Automatically generating such co-speech gestures is an important task in character animation and human-agent interaction, 
%for several reasons: First of all,
because a substantial fraction of our communication takes place through co-speech gestures 
\cite{mcneill1992hand, kendon2004gesture}. Furthermore, gesticulation has also been shown to enhance interactions with embodied agents \cite{bergmann2013virtual, luo2013examination},
e.g., to help with learning tasks \cite{bergmann2013virtual}, and
to lead to a higher sense of co-presence \cite{wu2014effects}.%, for example.

%The gesture-generation field has recently undergone a paradigm shift.
While early hand gesture-generation systems mainly relied on rule-based approaches \cite{cassell1994animated, kopp2004synthesizing, ng2010synchronized,marsella2013virtual}, data-driven gesture generation has become an important research area in recent years \cite{yoon2018robots, kucherenko2020gesticulator, yoon2020speech, ahuja2020no, ferstl2020understanding}. 
Both paradigms have advantages and disadvantages.
Rule-based systems produce gestures with clear communicative function, but lack diversity and require much manual effort to design. Data-driven systems, on the other hand, need less manual work and are more flexible, since they can generalise and generate new gestures on the fly. They may also scale better to large datasets. %However, most existing data-driven systems
%do not provide much control over communicative function, and
%generate gestures that have little semantic relation to speech content \cite{kucherenko2020genea}.
However, despite several attempts \cite{ahuja2020no,kucherenko2020gesticulator,yoon2020speech}, there have in our view been no convincing demonstrations of recent data-driven approaches consistently generating gestures with a clear semantic relation to the speech content.
%meaningful gestures generated from speech in a data-driven way.
For example, in terms of subjective gesture appropriateness for the speech, no system in the 2020 GENEA gesture-generation challenge \cite{kucherenko2020genea} surpassed a bottom line that simply paired the input speech audio with mismatched excerpts of training data motion, completely unrelated to the speech.

It would be desirable to develop approaches that combine the strengths of both paradigms, enabling systems to be built from data yet producing gestures that fulfil a communicative function together with the speech.
This has led us to investigate whether the communicative attributes of gesture can be modelled directly using recent data-driven methods.

The goal of this paper is to analyse to what extent modern deep-learning approaches are able to predict important communicative properties of hand gestures from the co-occurring speech.
As such, this work should \emph{not} be read as a machine-learning paper, since our focus is not to propose new architectures or advance the numerical performance on some pre-existing benchmark, nor as a gesture-generation paper, since no gesture synthesis is performed.
Instead, this is intended as a work on gesture analysis that studies the predictability of important gesture properties.
%and establishes new results that apply to a broad range of different gesture-generation systems.
%Our focus is not to propose specific network architectures nor to advance the numerical performance on some pre-existing benchmark, but to study the fundamental question of what aspects of gesticulation that feasibly can be predicted from the information accessible to automated systems.
Apart from being an interesting question in its own right,
%(and providing a kind of lower bound on the consistency/predictability of human gesticulation),
developing the ability to predict semantic aspects of gesticulation is a key element in driving future gesture-generation systems \cite{kucherenko2021speech2properties2gestures} to produce more meaningful and appropriate gesticulation.
%for driving future a gesture generation towards generating more meaningful and appropriate gestures in the future.
%This work can therefore be seen as a continuation of recent efforts \cite{ferstl2020understanding, saund2021cmcf, yunus2020sequence} towards imbuing systems for continuous, data-driven gesture generation with greater control over communicative function, with our new results laying down a path to reach that goal.
This work can therefore be seen as a continuation of recent efforts \cite{ferstl2020understanding, saund2021cmcf, yunus2020sequence} towards imbuing data-driven systems with greater control over communicative function.
%, and our findings suggest a path toward continuous generation of meaningful gestures.
%the results we establish lay down a path toward data-driven generation of meaningful, continuous gestures.
%gesture-generation approaches that produce meaningful gestures.
%are applicable to a broad range of different gesture-generation approaches.

%It should be noted that our goal is \emph{not} to create a full gesture-generation system -- indeed, no gestures are synthesised in this work. Instead, we establish new and important results on the predictability of meaningful gestures and their properties, results that apply to a broad range of different gesture-generation systems.

%In this paper, we analyse which gesture properties can be predicted from speech signal using deep learning.
The specific contributions of our work are:
\begin{itemize}
%\item We perform a study of different gesture-property prediction architectures, settling on a convolutional neural network with dilated convolutions.
%\item We propose to model binary gesture-property prediction by a convolutional neural network with dilated convolutions.
\item We conduct extensive gesture-property prediction experiments on a direction-giving dataset with a high fraction of representational gestures, for which gesture properties have been extensively hand-annotated. Specifically, we predict 13 distinct property labels -- 8 relating to communicative function -- which is significantly more than any prior work.
\item We analyse which modalities of speech -- audio and/or text -- are useful for predicting which gesture properties.
%and find that text helps to predict gesture category and semantics, while audio helps with predicting gesture phase
\item We investigate how individual or general different gesture properties are, by experimenting with gesture-property prediction for both known and previously unseen speakers.%seen versus previously unseen speakers.
%are and find that predicting gesture properties for novel speakers is more challenging, but still possible
%\item here might be more stuff ...
\end{itemize}
Despite the highly individual and stochastic nature of gestures, we find that numerous gesture properties can be predicted from speech% substantially better than chance
, both for speakers inside and outside the training data. % with $\text{F}_1$ scores around 60\%, including whether or not to gesture, gesture category, semantics, and gesture phase. Predicting gesture properties for speakers outside the training data is more challenging, but still possible.
We also find that speech text and audio differ in their uses, where time-aligned text enables predicting gesture category and semantics, while prosodic audio features help predict gesture phase.
%We also find that speech text and audio fulfil different functions for gesture property prediction, where text enables predicting gesture category and semantics, while audio helps predict gesture phase.
%We are in the process of releasing our code and anonymised processed data to improve reproducibility.
More information, including dataset and code, will be released on our project page at: \href{https://svito-zar.github.io/speech2properties2gestures/}{svito-zar.github.io/speech2properties2gestures}\ .
%
%To ease the reproducibility of our results,
%We will publicly release our code and anonymised data to simplify reproducing our results. 

%The remainder of this paper is laid out as follows:

\section{Related work}
\label{sec:related}

%In this work, we investigate the predictability of several different, concurrent properties of human gesticulation, based on human speech audio and text transcriptions.
Since this paper considers the predictability of different properties of human gesticulation from multimodal representations of speech, our review of related work covers two aspects: first the prediction of various gesture properties, and then the use and combination of speech modalities for gesture generation. 
%on the use of multiple speech modalities in gesture generation, and how they may be combined.
%the use of different speech modalities in deep learning for gestures.%, and how to combine the modalities.
%combining multiple modalities in machine learning systems.
%
%\subsection{Predicting gesture properties}
In general, the predictability of gesture properties has not been extensively studied, and most current gesture-generation systems do not integrate explicit gesture-property prediction, but there is nonetheless some prior work on predicting various gesture properties from speech.
%, which is surveyed below.% with the goal of enhancing automated gesture generation.
%, although the goal of that prediction need not always be for use in gesture generation.

\subsection{Gesture presence/absence prediction}
\citet{ferstl2021expressgesture} used a statistical method based on speech prosody peaks to predict where a gesture should be placed. They set the timing so that gesture strokes were 55\% complete at the pitch peak. \citet{yunus2019gesture} predicted gesture presence and timing based on speech audio using a recurrent neural network (RNN). In this work, we likewise explore using a neural network for this, but we use a convolutional neural network (CNN) instead of an RNN and consider a more extensive set of gesture properties.

\subsection{Gesture lexeme prediction}
Many gesture synthesis approaches predict gesture lexemes, or tags, that encapsulate both gesture form and semantics.  For example, a {\em cup} or {\em conduit} gesture involves a curved handshape, with the palm facing up and a forward motion of the hand from the speaker outward (gesture form) and is used to indicate an offering or conveyance (semantics). Systems of this type include  \cite{cassell2001beat, lee2006nonverbal, lhommet2013gesture, marsella2013virtual, kappagantula2020automatic, chiu2015predicting}. Some of these were rule-based, and predicted gesture semantics from input text based on a set of rules \cite{cassell2001beat,lee2006nonverbal, lhommet2013gesture}. Other research applied statistical methods to learn probabilistic mappings from semantic concepts to gestures \cite{kipp2005gesture, ishi2018speech}. Later, deep learning was applied to predict a fixed set of semantic gestures based on audio, text, and part-of-speech tags \cite{chiu2015predicting}. 
Our work, in contrast, does not consider a codified set of lexemes and instead predicts gesture properties that captures different elements of semantics, such as gesture categories and semantic gesture features.

%Cerebella \cite{lhommet2013gesture} is a system that automatically derives communicative functions from the text and audio of an utterance using a rule-based system. Communicative functions are then mapped to a multimodal behavior defined using the Behavior Markup Language \cite{kopp2006towards}.

%\citet{kappagantula2020automatic} presented a system capable of automatically performing deictic gestures for an animated pedagogical agent. The system takes the audio and text of the speech as inputs and triggers pre-animated deictic gestures based on keyword spotting. We are not aware of any machine-learning systems that can predict hand gestures' semantic meaning from speech.

\subsection{Gesture kinematics prediction}
\citet{ferstl2020understanding} considered predicting kinematic gesture properties (specifically velocity, initial acceleration, gesture size, arm swivel, and hand opening) from speech. They trained multiple recurrent neural networks to predict these gesture parameters from the speech audio signal, and found that some parameters, such as path length, were predicted more accurately than others, for example velocity. Instead of kinematics, we consider the predictability of gesture properties related to gesture semantics and phase.

\subsection{Gesture phase and category prediction}
%A gesture motion is typically segmented into several different phases, and there has also been work attempting to predict the gesture phase from speech \cite{yunus2020sequence}. 
Kendon \cite{kendon2011gesticulation} defined the following gesture phases: preparation, hold, stroke, and retraction. All phases are optional except for the stroke, which is the expressive phase of the gesture. It has been shown that gesture stroke is strongly correlated with pitch accentuation in speech \cite{jannedy2005structuring,esteve2013prosodic}. Furthermore, McNeill \cite{mcneill1992hand} defined different gesture categories, or dimensions, such as deictic, iconic, and metaphoric (all related to the spoken message) gestures and beat gestures (which are more strongly related to speech prosody and rhythm).

This paper investigates how well gesture phases (as defined by Kendon), gesture semantic meaning, and gesture categories (as defined by McNeill) can be predicted from speech audio and text in a data-driven manner. The most similar prior work is due to Yunus et al.~\cite{yunus2019gesture, yunus2020sequence}, where a restricted set of gesture phase and category were predicted
based on acoustic features only. Our study differs in that we consider additional gesture properties and also study the effect of different speech modalities as input.

\subsection{Effect of the speech input modality}
Many data-driven systems have only considered a single speech modality -- either audio recordings or text transcriptions thereof -- as input to the gesture generation, e.g., \cite{neff2008gesture, bergmann2009GNetIc, yoon2018robots, kucherenko2021moving}. However, the field is now shifting to use both audio and text together \cite{chiu2015predicting,kucherenko2020gesticulator, yoon2020speech, ahuja2020no}.
%, since they are considered complementary for predicting both gesture content and stroke timing.
This is, among other things, based on recent ablation studies of end-to-end gesture-synthesis systems in
\cite{kucherenko2020gesticulator,yoon2020speech}, that
%\citet{kucherenko2020gesticulator} and \citet{yoon2020speech}, which
%did independent ablation studies
compared gesture generation models which used only one modality against models using both. These studies found that using both speech modalities (audio and text) improved the synthesised gestures.
This paper delves further into the effects of the different input modalities, and addresses the question of which speech modalities are useful for predicting particular properties of human gesticulation.
%In this paper, we therefore consider \emph{speech} to be a multimodal signal that can be represented using either \emph{audio}, \emph{text}, or (most commonly) both.

%In our view, there have been no convincing demonstrations of meaningful gestures generated from speech in a data-driven way. For example, no system in the 2020 GENEA gesture-generation challenge \cite{kucherenko2020genea} surpassed a bottom line that simply played mismatched excerpts of training data motion, completely unrelated to the speech.

%This shortcoming led us to investigate whether the communicative attributes of gesture can be modelled directly.

\section{Data}

\subsection{Corpus}
There are two principal ways to obtain data for 3D gesture synthesis: optical motion capture \cite{lee2019talking, joo2019towards} and 3D pose estimation from videos \cite{yoon2020speech, ahuja2020no}. Among existing datasets, almost all are monologues, with only \cite{joo2019towards} involving interactions of more than one person.

Our present work aims at modelling iconic gestures, which are rare in all the previously cited datasets. Despite their important role in enabling meaningful gesticulation, these gestures only occur occasionally during social conversations. Hence we decided to focus on a %another
dataset that contains a large proportion of iconic gestures, the Bielefeld Speech and Gesture Alignment corpus (SaGA) \cite{lucking2013data}.
This is the largest and newest database we are aware of with detailed and accurate gesture-property annotations. Larger gesture databases exist, e.g., \cite{ahuja2020no}, but do not have the annotations necessary for our research.
We believe the SaGA dataset is sufficiently large for our purposes, since it has been previously used for generating iconic gestures \cite{bergmann2009GNetIc}, albeit based on information that cannot be extracted from speech.
%, but without analysing gesture-property predictability (as we do here).
%iconic gesture generation \cite{bergmann2009GNetIc} (but not to analyse the predictability).
%This dataset has previously been used for iconic gesture generation by \citet{bergmann2009GNetIc}.

The SaGA dataset contains a total of 280 minutes of recordings of 25 different participants speaking and gesturing to an interlocutor. Each recording lasted around 10 minutes, with durations ranging from  4 to 19 minutes. All recordings are in German. A key goal of SaGA was to capture a large number of iconic gestures. This was accomplished through a specific data collection procedure in which participants first saw a virtual reality bus tour and then described the route, and the prominent visual landmarks placed along that route, to another person. Both the navigation task and the landmarks provided natural visual grounding upon which iconic gestures are based. All participants followed the same route, thus
%, yet in their own ways, enabling semantic grounding and
maximising the degree of consistency between the recordings and simplifying the task of grounding gesture prediction in language by considering a tightly restricted semantic domain. Audio and video were recorded of each interaction \cite{lucking2013data} and every gesture was manually annotated according to a detailed labelling scheme. We use a subset of their annotation categories for our study, as described in \cref{sec:gestureproperties}.

%SaGA is the largest database we are aware of with detailed gesture-property annotations. Larger gesture databases exist, e.g., \cite{ahuja2020no}, but do not have property annotations necessary for the research we are conducting. The SaGA dataset is large enough since it has been previously used to build a model for iconic gesture generation \cite{bergmann2009GNetIc}.

\paragraph{Dataset partitioning}
%  we should use a comma everywhere 
Following previous works in gesture-synthesis research \cite{kucherenko2020gesticulator, alexanderson2020style, wu2021modeling}, the dataset was encoded at 20 fps. This resulted in 261,909 frames in total, out of which 127,581 frames were annotated as containing a gesture. For our research,
we replaced time-frames annotated as interlocutor speech with silence, in order to concentrate on the gesturing person's own speech.
%we masked with silence time-frames where the interlocutor was speaking (according to the annotation).
%We held out three recordings (numbers 7, 8, and 10) for testing and used the remaining 22 recordings for training and cross-validation. We did not use any of the test recordings in our experiments and left them as a test set for future research, so that future models can be evaluated without data leakage from the experiments reported here.
We used 22 out of 25 recordings for training and cross-validation. 
The remaining 3 recordings (numbers 7, 8 and 10) were held out for future research, so that future models can be evaluated without data leakage from the experiments reported here.

We used two different data partitions for cross-validation, to avoid tuning hyperparameters and evaluating on the exact same data splits. For choosing hyper-parameters, we performed classical 10-fold cross-validation. For evaluating the model, we use 20-fold cross-validation, set up such that every fold contains 5\% of the data from each of the 22 subjects in the recordings we consider. Training and validation sequences never overlapped.

\subsection{Speech modalities and their encoding}

We used two different speech modalities from the dataset, each of which is described below.

%\paragraph{Motions}
%We extracted 3D join angles of the actors from the dataset using FrankMocap [?]. The motion was further processed by ... Finally, we obtain all the motions in BVH format.

\paragraph{Text}
Each recording was transcribed in German. Transcriptions contain the written form of every word and its timing (onset and offset), but no punctuation or other sentence delimiters due to the spontaneous and continuous nature of the speech.

We experimented with two commonly used word embeddings for German: DistilBERT \cite{sanh2020distilbert}, which encodes each word together with context, and FastText  \cite{joulin2016fasttext}, which does not take context into account. FastText outperformed DistilBERT when predicting the semantics property (where the text modality has the most impact) and was hence chosen as the text embedding for our experiments.

%Text features were extracted by applying German FastText \cite{joulin2016fasttext} to the full sequence of words spoken by each the participant.
The FastText tokeniser produces one 300-dimensional feature vector (a.k.a.\ ``embedding'') per word-piece token. These were converted to a single feature vector per word by computing the arithmetic average of the feature vectors of all word pieces within that word.
When predicting gesture properties, each vector was supplemented with one extra number about word timing, namely
the time-difference from the word onset to the prediction target frame (negative for words starting before the target point and positive for future words).
Text-based gesture-generation commonly uses timing information \cite{ishi2018speech,yoon2018robots}, even though that information cannot be derived from text alone.

\paragraph{Audio}
We extracted the audio tracks from each video and converted them to mono waveforms with a 48 kHz sampling rate. We then used Parselmouth~\cite{jadoul2018introducing} to compute five prosodic features as the audio feature set of our experiments: voiced/unvoiced binary flag, log fundamental frequency (linearly interpolated in unvoiced regions), log energy, and the derivatives of the last two computed with finite differences.
%For each recording, we extracted the audio tracks from the interaction videos in the database and converted these into mono waveforms sampled at 48 kHz. From these audio signals, we used Parselmouth~\cite{jadoul2018introducing} to extract four prosodic features to use as the audio feature set in our experiment, namely pitch (F0), energy and their derivatives  computed with finite differences.
Such prosodic features are commonly used in speech emotion analysis as well as for gesture-property prediction, e.g., \cite{yunus2019gesture}. Specifically, we transformed pitch and intensity like in \cite{chiu2011train, kucherenko2021moving}: the pitch values were adjusted by taking $log(x+1)-4$ and setting negative values to zero, and the intensity values were adjusted by taking $log(x)-3$.
The audio features were first extracted at 200 fps
%, as is common in speech analysis, and were
and then resampled to 20 fps by averaging, to match the resolution of the gesture annotations.

We also experimented with using spectrograms instead of prosodic features, but found no difference between the two when predicting gesture phase (where the audio modality has the most impact). 
%, except that prosody is 
Prosodic features were chosen since they have the benefit that they are more anonymous, enabling us to release audio features.
%We use prosodic as it is anonymous and hence enables us to release audio features.

\begin{figure}
    \centering
    \includegraphics[width=0.49\textwidth]{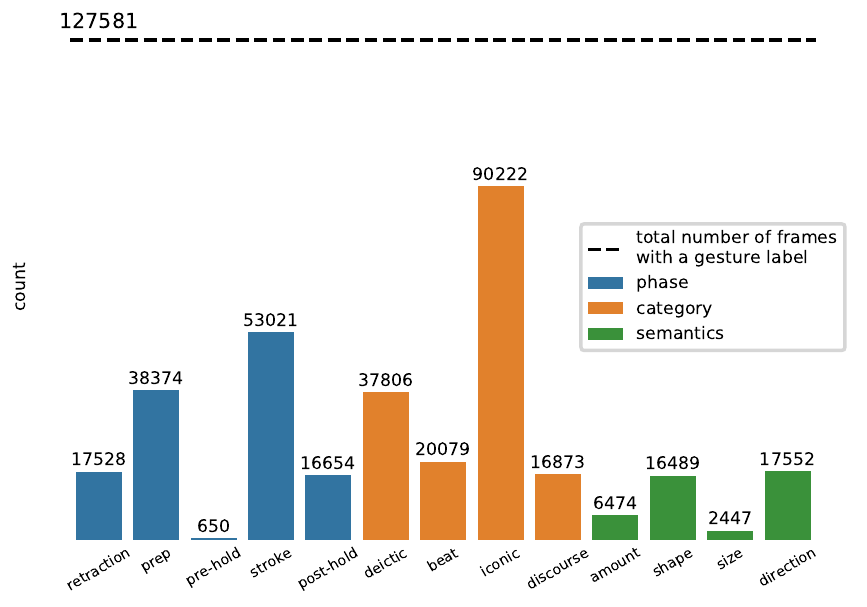}
    \caption{The frequency of each gesture-property label in the SaGA dataset. Note that frequencies may sum to more than  127,581 since most categories are not mutually exclusive.}
    \label{fig:label_stats}
    \Description{A bar plot containing the number of frames with each of the 13 possible gesture-property labels. The numbers are as follows. Phase labels: retraction: 17528, prep: 38374, pre-hold: 650, stroke: 53021, post-hold: 16654.
    Category labels: deictic: 37806, beat: 20079, iconic: 90222, discourse: 16873.
    Semantic labels: amount: 6474, shape: 16489, size: 2447, direction: 17552.}
    \vspace{-4mm}
\end{figure}

%Since our models considered a window of five frames of past and future context around each time frame when making predictions, the full receptive field of our prediction models was 11 frames, or 2.4 seconds, of audio.
%We also experimented with mel-spectrogram features, but these features may be less transferable between speakers and we did not observe any improvements in prediction performance.

\subsection{Gesture properties and their encoding}
\label{sec:gestureproperties}

%\paragraph{Property annotations}
The SaGA corpus contains detailed annotations of the properties of the gestures in the recordings. We made use of the following gesture properties in our experiments: \textit{R.G.Left Semantic},
\textit{R.G.Right Semantic},
%\textit{R.G.Left Practice},
\textit{R.G.Left Phrase}, \textit{R.G.Right Phrase}, \textit{R.G.Left.Phase}, and \textit{R.G.Right.Phase}. The \textit{Semantics} property indicates which semantic information is contained in the gesture. \textit{Phrase} indicates gesture category. \textit{Phases} are sub-units of gestures that indicate: if the hands are preparing to gesture, meaning is currently being conveyed, etc.   For details about the data collection and annotation scheme we refer the reader to \citet{lucking2010bielefeld} and \citet{bergmann2006verbal}. To simplify modelling, we merged the features for the left and right hand into a single feature using a per-frame logical OR. Each feature was encoded into a vector of binary values, which is one-hot for \textit{Phase} since phases are mutually exclusive.

%Since this work is about predicting gesture properties from speech text and audio features, the studies in this paper do not rely on the video recordings., none of the studies in this paper used the RGB video streams contained in the dataset in any way. All of the information relevant to our research questions is contained in the above-described speech material and annotations (which, of course, were created using the video recordings as one of the input sources).

\paragraph{Gesture-property representations}

We encoded gesture properties at a rate of twenty frames per second (20 fps). %from all 25 recordings in the dataset
As described in Section~\ref{ssec:model}, our system first predicts if a gesture is needed and then what kind of gesture it should be. For the latter gesture-property prediction task, we only consider time-frames where a gesture was present in the data, i.e., frames where any of the annotations we considered were present and nonzero. This amounted to 127k out of 261k total frames.
We list the gesture-property labels we considered and the number of frames they were present at in Figure~\ref{fig:label_stats}. As can be seen, most of the gesture-property labels only apply to a small fraction of the gesture-containing frames in the data.

\iffalse
\begin{table}
    \centering
    \begin{tabular}{@{}|l|l|l|l|l|@{}}
    \hline
        Ges.\ category & deictic & beat & iconic & discourse \\ \hline
        No.\ of frames & 9720 (29\%) & 4841 (14\%) & 24096 (72\%) & 4277 (13\%)\\ \hline  \hline \hline
        
        Ges.\ semantics & amount & shape & direction & size \\ \hline
        N. of frames & 1582 (5\%) & 4383 (13\%) & 4585 (14\%) & 650 (2\%) \\ \hline
    \end{tabular}
    \caption{Statistics for mutually not-exclusive gesture categories: semantics and categories. There are 127,581 frames in total. The percentages sum to more than 100\% since gesture categories are not mutually exclusive.}
    \label{tab:StatSemantPhrase}
    \vspace{-3mm}
\end{table}

\begin{table}
    \centering
    \begin{tabular}{@{}|l|lllll|@{}}
    \hline
        Ges.\ phase & prep & pre-hold & stroke & post-hold & rest \\ \hline
        %N. of frames & 10306 (30\%) & 184 (0.5\%) & 13668 (40\%) & 4096 (12\%) & 4951 (15\%) \\ \hline
        N. of frames & 10306 & 184 & 13668  & 4096 & 4951 \\ \hline
    \end{tabular}
    \caption{Statistics for gesture phase out of 33454 frames.}
    \label{tab:StatPhase}
    \vspace{-3mm}
\end{table}

\fi

We encoded the gesture properties as binary vectors. For this, we first created an ordering of the different labels relevant to each property. For example, for \textit{Gesture Category} we ordered the different possible labels as follows \{1: `deictic', 2: `beat', 3: `iconic', 4: `discourse'\}. A frame with Category annotation ``beat-iconic'' would then be encoded by the vector $[0, 1, 1, 0]^T$.
As the example shows, gesture categories are not mutually exclusive, and several labels can be present simultaneously. The same applies to gesture semantics labels. Gesture phase, on the contrary, is exclusive -- only one label can be applicable at a time -- and we take this mutual exclusivity of gesture phases into account during modelling and evaluation.
%Hence those properties are treated differently during modelling and evaluation.

Note that the work in this paper does not make use of the videos captured during the SaGA corpus recordings, only transcriptions, gesture annotations, and anonymous audio features (prosody) derived from those recordings.
We will release the extended and anonymized version of the SaGA dataset at our project page: \href{https://svito-zar.github.io/speech2properties2gestures/}{svito-zar.github.io/speech2properties2gestures}\ .

\section{Experimental setup}

% For us - terminology used:
% each of the gesture properties (e.g. phase or "semantics") would have a set of labels that may or may not be mutually exclusive. Finally, a label is either absent or present in a given frame.

This section describes the experimental setup for our experiments on predicting gesture properties %(such as gesture category and phase) 
from speech text and audio.

\begin{figure}
\centering
  \includegraphics[width=0.49\textwidth]{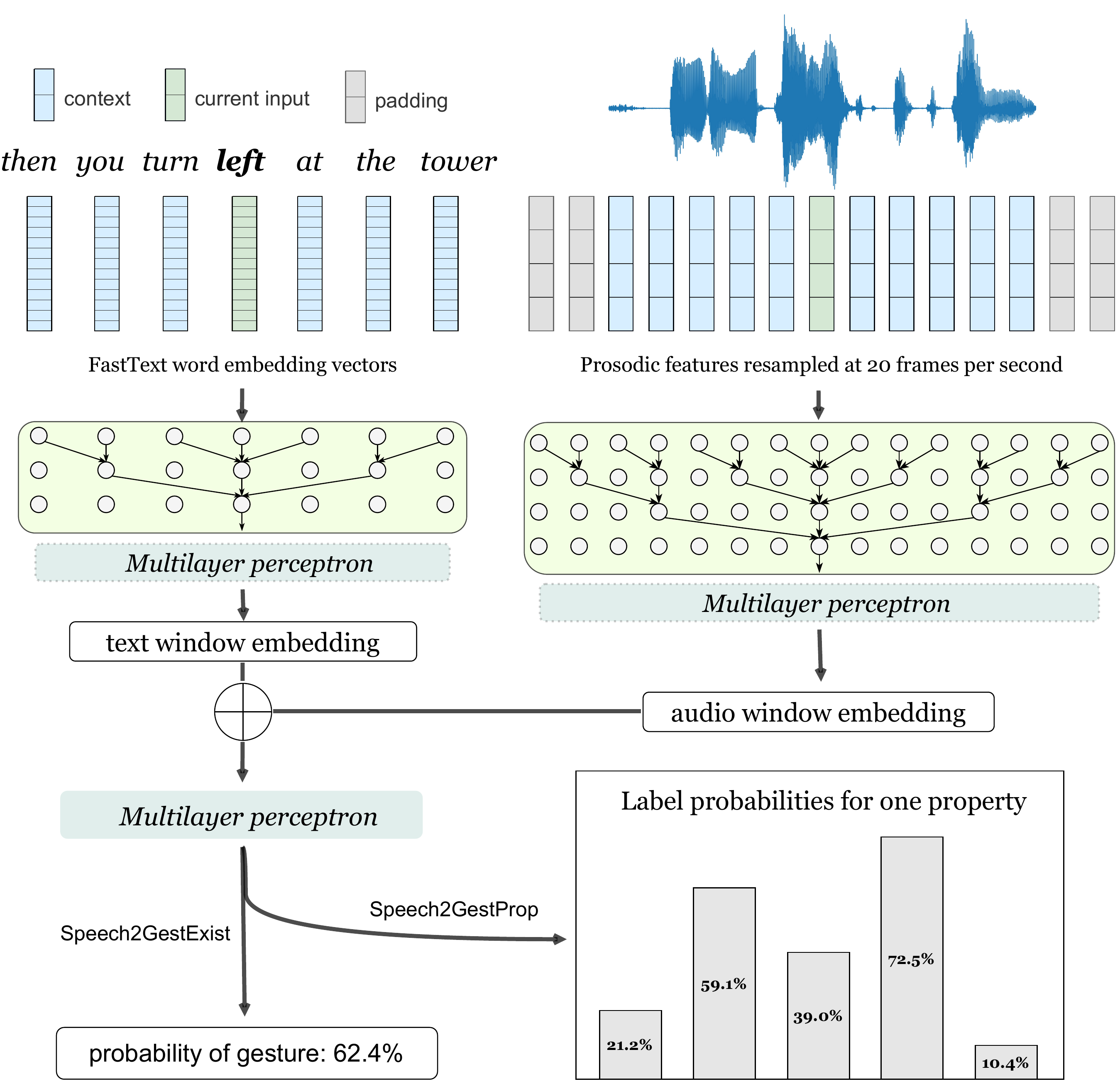}
  \caption{The shared multimodal architecture of our two networks. First, the two modalities are independently encoded using dilated temporal CNNs, using zero-padding as necessary. Then, the two encodings are concatenated and fed into an MLP decoder, which returns the final output.}%  Overview of the proposed framework. We use speech text and audio to predict several binary gesture properties, such as gesture type.}
  \Description{An architecture figure for the multimodal neural network. On the top of the figure, the inputs are shown side-by-side: text, with the example sentence "then you turn left at the tower"; and audio, depicted with a waveform. The two inputs are converted to a sequence of vectors, centred around the middle frame. The vectors are FastText word embeddings for text (per-word), and prosodic features for audio (at 20 FPS). The prediction is then generated as described in the caption. It may be the probability of gesture, or the label probabilities for one property, for the middle frame of the input.}
  \label{fig:model}
\end{figure}

\subsection{Problem formulation}

We frame the problem of gesture-property prediction as follows: given a sequence of speech features $\boldsymbol{s} = [s_t]_{t=1:T}$ %extracted from segments (frames) of speech audio at regular intervals $t$
the task is to generate a sequence of corresponding binary gesture properties $\boldsymbol{\hat{p}} = [\hat{p}_t]_{t=1:T}$. Here, $t=1:T$ denotes indexing into a sequence of vectors for integer $t$ in $1$ to $T$. 
Each speech segment $s_t$ is represented by several different features, specifically acoustic features (e.g., prosody), semantic features (e.g., word embeddings), or both.

\subsection{Gesture-property prediction model}
\label{ssec:model}
Our gesture-property prediction model %proposed in this paper 
 consists of two components that take speech audio and text as input: % illustrated in Figure \ref{fig:teaser}:
\emph{Speech2Gest\-Exist}, %which takes speech (both text and audio) as input and returns 
which predicts the probability of making a gesture%(similar to \cite{yunus2019gesture})
, and \emph{Speech2\-Gest\-Prop}, which %takes speech as input and
predicts the probabilities of different labels for a given gesture property. %a set of binary gesture properties, such as gesture type, gesture phase, etc.
Such hierarchical models have been successful on other sequence-prediction tasks such as text-to-speech intonation generation, where first predicting the presence or absence of voicing, and then predicting voicing frequency, worked better than predicting the two aspects jointly at once \cite{wang2018autoregressive}.

\paragraph{Detailed model specification}
\label{ssec:modeldetails}
We implemented the \emph{Speech2GestProp} and \emph{Speech2GestExist} components using the same architecture based on dilated convolutions \cite{yu2016multi} for information aggregation along the time dimension. We have chosen convolutions instead of recurrency because no long-term memory is needed for this task, since what was said one minute ago is irrelevant for the present gesture. 
Dilated CNNs \cite{yu2016multi} are a widely-used neural-network architecture for sequence modelling, used in WaveNet \cite{vandenoord2016wavenet} and WaveGlow \cite{prenger2019waveglow}, and 
%which has been 
recently also adapted to human motion modelling \cite{hou2021causal}.
%Following the developments in \cite{vandenoord2016wavenet, prenger2019waveglow}, we evaluated the effect of residual and skip connections, different activation functions and weight initialisation schemes. We also experimented with no dilation and with a simple MLP \cite{rosenblatt1958perceptron}. The experimental results are publicly available on FigShare: \href{https://doi.org/10.6084/m9.figshare.15134295}{doi.org/10.6084/m9.figshare.15134295}. Seeing that neither of these modifications provided any benefits, we use the dilated convolution architecture depicted in \cref{fig:model} for all other experiments.

%Using dilated convolution is an efficient way to model temporal sequences, and removing dilation made the model perform worst. On top of that, we also experimented with adding residual connections, removing convolution and changing the weight initialisation scheme. Neither of these modifications provided any benefit, and hence we use a convolutional neural network with dilated convolutions,  Our model architecture is illustrated in Figure~\ref{fig:model}.

The model inputs are sequences of audio frames and transcribed spoken words in a sliding window centred on the current time frame. Based on findings regarding the temporal synchrony between speech and gesture 
%speech-gesture temporal synchrony
\cite{loehr2012temporal, pouwQuantifying}, we consider the current, three past, and three future word-token feature vectors %\footnote{In our experiments, these features were timing information and DistilBERT contextual word embeddings, which depend on the entire sentence each word belongs to, so our system predictions can be informed also by words outside this 7-token window.}
and the current and twenty past and twenty future audio frames (i.e., 1 s to either side). %\emph{Speech2GestExist} returns the estimated probability of gesture existence, while \\ \emph{Speech2GestProp} returns a vector of probabilities, one for each relevant gesture property, all at time frames in the middle of the input window.
By sliding these windows over the input speech-feature sequences, we can make predictions for the selected gesture properties frame by frame for all times $t$ in the sequence. (For this paper, we only considered frames sufficiently far from sequence edges for all model inputs to be well defined, to avoid edge effects.)
This setup makes use of future speech, which is standard in gesture generation and rarely considered a limitation since most applications do not depend on live speech.
For example, the utterance-based TTS systems used by many social robots and virtual agents require the entire utterance text to be available before audio synthesis can begin.

As illustrated in Figure \ref{fig:model}, speech audio and text are first encoded into the intermediate \textit{text window embedding} and \textit{audio window embedding} representations using two separate neural networks, each of them containing several layers of dilated convolution. The two embeddings are then concatenated and passed into a simple fully-connected neural network (MLP).
%, a.k.a.\ a Multi-Layer Perceptron (MLP) \cite{rosenblatt1958perceptron}.
At the final layer, we map the values onto the unit interval $[0,1]$, since the output should indicate the probability that each relevant gesture property is present. For that, a sigmoid output nonlinearity is applied to \textit{Gesture Category} and \textit{Gesture Semantics} and a softmax output nonlinearity is applied to the \textit{Gesture Phase} outputs. The softmax is used since different phase labels, unlike the other property categories, are mutually exclusive. From a probabilistic perspective, the use of a sigmoid for each binary property corresponds to the assumption that each property is statistically independent of the others, given the input features. This is a common modelling assumption for binary variables that are not mutually exclusive.

%Text features were \rn{extracted using} DistilBERT \cite{sanh2020distilbert} ( simplified and compressed version of BERT \cite{devlin2018bert}) as implemented by HuggingFace \cite{wolf-etal-2020-transformers}.  For audio features, we used four prosodic features: pitch (F0), energy, and their derivatives, commonly used in speech emotion analysis as well as for gesture properties prediction \cite{yunus2019gesture}.

Since any given gesture property is present in just a
fraction of the time frames, any gesture-property predictor training data will be highly imbalanced.
To mitigate this, we experimented with upsampling underrepresented classes to balance the data and also considered several different loss functions:
not only the standard \emph{cross-entropy loss}, $CE(p_t) = -log(p_t)$ (where $p_t$ is the model probability of the correct class at time $t$), but also the \emph{focal loss} \cite{lin2017focal}, developed to address the rarity of positive labels in common datasets, and a class-balancing version of the focal loss from \cite{cui2019class}.
The results are reported in Section~\ref{ssec:mlsetup}.
%\begin{enumerate}
%\item Standard cross-entropy loss, $CE(p_t) = -log(p_t)$, where $p_t$ stands for the probability the model assigned to the correct class at time $t$.
%\item \emph{Focal loss} \cite{lin2017focal} was developed to address the rarity of positive labels in common datasets. It is defined by \begin{align} FL(p_t) = -(1-p_t)^\gamma log(p_t) \text{,}\end{align} where $\gamma$ is a tunable parameter.
%\item A class-balancing version of the focal loss due to \citet{cui2019class}, defined by \begin{align}CB(p_t, y) = \frac{(1-\beta)}{(1-\beta^{n_y})} FL(p_t, y)\text{,}\end{align} where $n_y$ is the total number of examples of the ground-truth class $y$ and $\beta$ is a parameter which is usually between 0.5 and 1. As seen from the formula, this loss function weighs different classes differently based on how common they are in the data.
%\end{enumerate}
Each loss function is aggregated for sequences and minibatches by summing over constituent frames.

\paragraph{Hyperparameters}

For each experiment and each model in Section~\ref{sec:experiments}, we conducted a separate hyperparameter search using random search \cite{bergstra2012random}. Each random search consisted of 50 runs. For each run, we randomly sampled all the key hyperparameters over a predefined range for each value and trained the model for a fixed number of epochs dependent on the task. We found no significant difference in the validation scores in the latter half of training, therefore no early stopping was used and the weights from the final epoch were used. During the hyperparameters search we varied: hidden dimensionality, number of layers, kernel size, dropout, and output embedding dimensionality for each encoder; hidden dimensionality, number of layers, and dropout for the decoder; learning rate, batch size, and other optimisation parameters.

We selected the best hyperparameters based on the average Macro $\text{F}_1$ \cite{yang1999re} score of 10-fold cross-validation, and used these settings to compute the results reported in Section~\ref{sec:experiments}. Hyperparameters for all models in the paper are publicly available on FigShare: \href{https://doi.org/10.6084/m9.figshare.15134076}{doi.org/10.6084/m9.figshare.15134076}.

\vspace{-1mm}
\subsection{Baseline systems}
\label{ssec:baselines}
For the majority of the properties we predict, no previous baseline systems or benchmark performance exist.
Instead, our main starting point for baselining is the finding from \cite{kucherenko2020genea} that no gesture-generation system beat a mismatched bottom line that paired speech with unrelated training-data motion.
%With this work we have no intention to beat the state-of-the-art in gesture-property modelling, we rather investigate what can be predicted and under which conditions.
%baseline systems exist for direct comparison. In the absence of such baselines, we
%That is why instead of comparing with other machine-learning models, w
Inspired by this, we create and compare against a number of simple bottom-line systems that similarly have no dependence on the input speech. These include two constant-output systems (\emph{AlwaysZero} and \emph{AlwaysOne}), and two systems based on random output, either uniformly random (system \emph{UniformRandom}) or random draws with the same distribution as the a-priori class abundances in the training data (system \emph{InformedRandom}).
%We consider all of these to be bottom-line systems, since none of them leverage any input features for improved prediction -- they have no dependence on the speech.
Any system can be said to be \emph{better than chance} if it surpasses all four of these bottom lines. Moreover, any time that happens, we say that the corresponding property is \emph{predictable} from the given input features.
(This is very different from being perfectly predictable, which arguably is an unrealistic goal for problems that involve human behaviour.)
%Even a minor improvement over chance predictability could add important communicative value to generated gestures, since current state-of-the-art gesture generation is no more appropriate than random gestures \cite{kucherenko2020genea}.
%enabling systems to replace ``low meaning'' gestures with semantically appropriate ones even a fraction of the time, could add important communicative value.

%Note that our definition of ``predictable'' is very different from being \emph{perfectly predictable}:

\vspace{-1mm}
\subsection{Evaluation metrics}
It is well known that standard classification accuracy (one minus the error rate) does not capture overall system performance well when the data is highly unbalanced, since it may then be possible to achieve high accuracy by always predicting the majority class, regardless of the input features of the given instance. 
Instead, we use the $\text{F}_1$ score as our main performance indicator. This measure is the harmonic mean of precision and recall, and is a popular evaluation measure for classification of unbalanced classes. More specifically, we use the Macro $\text{F}_1$ score \cite{opitz2019macro}, 
which is simply the arithmetic average of $\text{F}_1$ scores for all possible, mutually exclusive classes $c$: $\text{Macro F}_1 = \frac{1}{C}  \sum_{c=1}^C \text{F}_1 (c)$.

Note that since phase labels are mutually exclusive, while other gesture-property labels are not, phase is evaluated differently. For gesture categories and semantics we calculate separate Macro $\text{F}_1$ scores for each label, since they are not mutually exclusive and are treated as independent. For the gesture phase, on the contrary, we evaluate only the $\text{F}_1$ scores for each label, and not the Macro $\text{F}_1$ score, which averages over all possible labels.
%and one Macro $\text{F}_1$ score as an average of the binary $\text{F}_1$ scores for all the gesture phase labels. 

To get a better understanding of generalisation ability on our limited dataset, we used cross-validation. For each of our experiments, we report the mean and standard deviation of of the selected performance measure across 20 cross-validation folds.
These folds were set up such that every fold contained 5\% of the data from each of the 22 people in the recordings we considered. This means that the cross-validation quantifies \emph{within-person generalisation performance}, although we also looked at \emph{across-person generalisation} by holding out one individual at a time (see Section~\ref{sec:generalization_study}).

\section{Results and Discussion}
\label{sec:experiments}

% Better-than-chance formatting
%\newcommand{\highlight}[1]{\textbf{\textcolor{Maroon}{#1}}}
\newcommand{\highlight}[1]{\textbf{\textcolor{OliveGreen}{#1}}}
\newcommand{\showissue}[1]{\textbf{\textcolor{Maroon}{#1}}}

We conducted several experiments, first comparing different performance metrics, and then evaluating 1) how well we can predict gesture presence, 2) which modalities are essential for predicting which gesture properties, 3) how well predictions generalise to new speakers, and more.
In this section, we report and discuss the results of these experiments.

% That is why as our main metric, we use the $\text{F}_1$ score, which is handling class imbalance well. We follow a common approach and use Macro $\text{F}_1$ score \cite{opitz2019macro}, which is an average of $\text{F}_1$ scores for all the possible classes.

%\paragraph{Default settings}
In each experiment, we vary one aspect while keeping everything else the same. Our default settings %, from which we change one aspect at a time, 
are:
\begin{itemize}
\item using both speech modalities, instead of only audio or text;
\item evaluating generalisation within known speakers, instead of generalisation to new speakers;
\item training individual models for each gesture property, instead of training a single model of all properties simultaneously.
\end{itemize}

\subsection{Comparison of evaluation metrics}
In order to put the evaluation metric used into context, Table~\ref{tab:Detailed_metrics} reports the accuracy, precision, recall, $\text{F}_1$ and Macro $\text{F}_1$ scores for predicting the presence/absence of (as an example) the gesture semantics property label ``shape''.

Overall, Macro $\text{F}_1$ is the most preferable evaluation metric. Accuracy is misleading because it can be very high for primitive baselines (such as AlwaysZero) simply because one class is dominant over the other. 
%Macro $\text{F}_1$ is a preferable evaluation metric. Accuracy is misleading because accuracy can be very high for very primitive systems (such as for AlwaysZero) just because one class is dominant over the other. 
Using only precision or recall is not sufficient, as each focuses only on either false negatives or false positives. As the $\text{F}_1$ score for label presence is the harmonic mean between precision and recall, %. As a harmonic mean, 
it tends to be closer to the lower of the two values (see, e.g., UniformRandom). However, the $\text{F}_1$ score is not symmetric and strongly focuses on true positives. The Macro $\text{F}_1$ score is computed as the arithmetic average between the $\text{F}_1$ scores for label presence ($\text{F}_1$ for 1) and for label absence ($\text{F}_1$ for 0). This metric has the added advantage that chance performance is 50\% even for imbalanced data, and it is the metric we choose to report in the rest of this section.
The maximum achievable score, on the other hand, is likely significantly below 100\%, since human gesticulation is highly stochastic.

%Interestingly, our model achieves much higher precision than recall and a much lower $\text{F}_1$ score for one than for zero. The accuracy and recall are more similar to AlwaysZero than Always\-One, suggesting that the model may be overpredicting the majority class, as is typical of cross-entropy-based classifiers on imbalanced data.
%of thresholding-based classifiers.

 % transposed table
\begin{table}
    \centering
    % Coloured rows
    %\rowcolors{2}{}{lightgray} 
    % Fit the whole page
    \resizebox{0.475\textwidth}{!}{
    % Taller lines
    \renewcommand{\arraystretch}{1.5}
\begin{tabular}{
    % Magic for aligning on plus-minus symbol, see property_prediction.text
    % \kern fixes an issue with grey colouring being too wide
    @{\kern\tabcolsep} r *{18}{r@{\hspace{\tabcolsep}}c@{\hspace{\tabcolsep}}l}@{\kern\tabcolse}
}
\specialrule{\heavyrulewidth}{0pt}{0pt}       & \multicolumn{3}{c}{\textbf{Accuracy}} & \multicolumn{3}{c}{\textbf{Precision}} & \multicolumn{3}{c}{\textbf{Recall}} & \multicolumn{3}{c}{\textbf{$\text{F}_1$ for 1}} & \multicolumn{3}{c}{\textbf{$\text{F}_1$ for 0}} & \multicolumn{3}{c}{\textbf{Macro $\text{F}_1$}} \\

\specialrule{\lightrulewidth}{0pt}{0pt}
AlwaysZero  & \showissue{87\%} & \showissue{$\pm$} & \showissue{10\% }& 0\% & $\pm$ & \hphantom{0}0\%   & 0\% & $\pm$ & \hphantom{0}0\%   & 0\% & $\pm$ & \hphantom{0}0\% & \showissue{92\%} & \showissue{$\pm$} & \showissue{2\%}           & 46\% & $\pm$ & 1\%           \\ 

\rowcolor{lightgray}
\textit{AlwaysOne}   & 13\% & $\pm$ & 5\% & 13\% & $\pm$ & 5\% & \showissue{100\%} & \showissue{$\pm$} & \hphantom{0}\showissue{0\%} & 22\% & $\pm$ & 8\%          & 0\% & $\pm$ & 0\% & 11\% & $\pm$ & 4\%           \\ 

\rowcolor{white}
\textit{UniformRandom} & 50\% & $\pm$ & \hphantom{0}1\%  & 13\% & $\pm$ & 5\% & 50\% & $\pm$ & \hphantom{0}2\%  & 20\% & $\pm$ & 6\%          & 64\% & $\pm$ & 5\%          & 42\% & $\pm$ & 3\%           \\ 

\rowcolor{lightgray}
\textit{InformedRandom}  & \showissue{77\%} & \showissue{$\pm$} & \hphantom{0}\showissue{4\%}  & 12\% & $\pm$ & 5\% & 12\% & $\pm$ & \hphantom{0}1\%  & 12\% & $\pm$ & \hphantom{0}3\%           & \showissue{86\%} & \showissue{$\pm$} & \showissue{2\%}           & 49\% & $\pm$ & 1\%           \\ 

\specialrule{\lightrulewidth}{0pt}{0pt}

our result       & 86\% & $\pm$ & \hphantom{0}4\%  & 44\% & $\pm$ & 12\% & 35\% & $\pm$ & 9\% & 39\% & $\pm$ & 9\%    & 92\% & $\pm$ & 3\%           & 67\% & $\pm$ & 5\%      \\  

\specialrule{\heavyrulewidth}{0pt}{0pt}    \end{tabular}}

    \caption{A comparison between various evaluation metrics for gesture property prediction for the gesture semantic property ``shape''. The baselines are \textit{italicised}; ``our result'' refers to our multimodal dilated CNN (BothModalities). \showissue{Red colour} highlights issues with the associated metrics. } \label{tab:Detailed_metrics}
\end{table}

\subsection{Predicting gesture presence and timing}

\begin{figure}
\centering
  \includegraphics[width=0.4\textwidth]{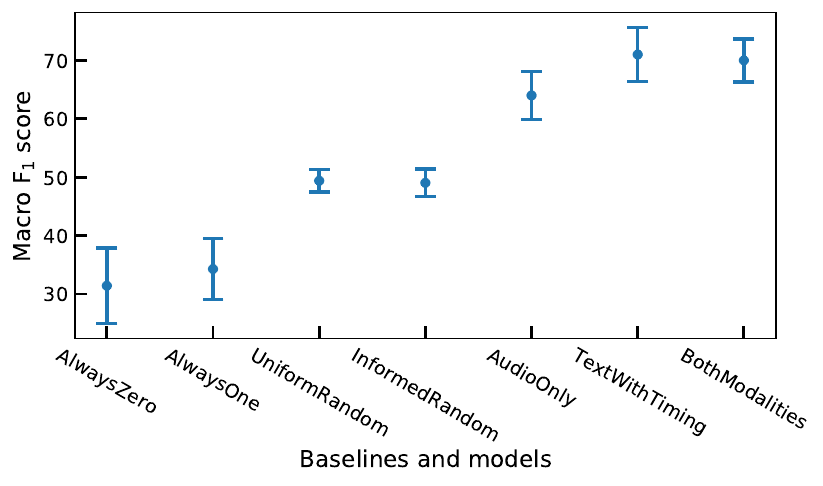}
  \caption{Macro $\text{F}_1$ scores (means and standard deviations) for gesture presence prediction. Gesture presence is seen to be predictable regardless of the input modality used.}
  \label{fig:Speech2GestExist}
  \Description{A point chart showing the Macro F1 scores for 4 baselines and 3 models as follows. AlwaysZero - 31.41\%, AlwaysOne - 34.28\%, NaiveRandom - 49.39\%, Informed Random - 49.03\%,  AudioBased - 64\%, TextBased - 71\%, FullModel - 70\%.}
\end{figure}

The first question we considered was whether or not it is possible to predict when to make a gesture
%First of all, we studied . First, we validated that whether or not a given time frame contains a gesture
%the presence or absence of gesture for a given time frame is possible to predict
%More specifically, we studied to what extent we were able to predict
(i.e., predict the presence or absence of a gesture from the speech features in our dataset).
The best model found by our hyperparameter search achieved a $70\% \pm 3.7\%$ Macro $\text{F}_1$ score on this binary classification task. This is better than %previous work \cite{yunus2019gesture}, which achieved $\text{F}_1$ scores of 46.3\% and 72.7\% (on a different dataset than ours), resulting in a Macro $\text{F}_1$ score of 59.5\%, and also better than
chance (see Figure~\ref{fig:Speech2GestExist}) and agrees with results from previous work on another dataset in a different language (English) \cite{yunus2019gesture}. 

\subsection{Experiments on machine-learning setup}
\label{ssec:mlsetup}
%In this section, we report on our ablation experiments.
We experimented with two different approaches for dealing with data imbalance, namely upsampling uncommon classes and special loss functions, as described in Section~\ref{ssec:modeldetails}.
Our experiments found no benefits to either upsampling or the special loss functions in terms of Macro $\text{F}_1$ score: the results were slightly better for some features and slightly worse for the others, but there was no major difference. We therefore use the conventional cross-entropy loss without any upsampling for all other experiments in this paper.
%, therefore, do not apply those class-balancing techniques in the subsequent experiments. 

We also experimented with training a single model to predict all the gesture properties at once, versus training individual models for each gesture property. We found that using individual models has higher macro $\text{F}_1$ score, possibly due to different tasks benefiting from different hyperparameter choices, hence we model each gesture property individually in the rest of our experiments.

\subsection{Evaluating text and audio contributions}
\begin{table*}[t]
\centering
%\rowcolors{3}{}{lightgray} % Coloured rows

% Special commands for "extra bold" fonts for some words in the table caption (which is entirely bold by default)
% source: https://tex.stackexchange.com/a/445173
\newcommand{\fbseries}{\unskip\setBold\aftergroup\unsetBold\aftergroup\ignorespaces}
\makeatletter
\newcommand{\setBoldness}[1]{\def\fake@bold{#1}}
\makeatother

% This command only affects the "extra bold" style in the caption
\setBoldness{0.6}
% Make the table fit the whole page
\resizebox{\textwidth}{!}{
% Taller lines
\renewcommand{\arraystretch}{1.5}

%One right-aligned column followed by 13*3=39 columns (e.g. deictic is 3 columns because I align on the plus-minus by putting it into a separate middle column). The column triplets are right-center-left aligned.
\begin{tabular}{@{\kern\tabcolsep}r *{39}{r@{\hspace{\tabcolsep}}c@{\hspace{\tabcolsep}}l}@{\kern\tabcolsep}}

\toprule
& \multicolumn{12}{c}{\textbf{gesture category} [Macro $\text{F}_1$]} & \multicolumn{12}{c}{\textbf{gesture semantics} [Macro $\text{F}_1$]} & \multicolumn{15}{c}{\textbf{gesture phase} [$\text{F}_1$]} \\

\cmidrule(lr){2-13}
\cmidrule(lr){14-25}
\cmidrule(lr){26-40}

label & \multicolumn{3}{c}{deictic} & \multicolumn{3}{c}{beat} & \multicolumn{3}{c}{iconic} & \multicolumn{3}{c}{discourse} & \multicolumn{3}{c}{amount} & \multicolumn{3}{c}{shape} & \multicolumn{3}{c}{direction} & \multicolumn{3}{c}{size} & \multicolumn{3}{c}{pre-hold} & \multicolumn{3}{c}{post-hold} & \multicolumn{3}{c}{stroke} & \multicolumn{3}{c}{retraction} & \multicolumn{3}{c}{preparation} \\

relative frequency & \multicolumn{3}{c}{29.05\%} & \multicolumn{3}{c}{14.47\%} & \multicolumn{3}{c}{72.03\%} & \multicolumn{3}{c}{12.78\%} & \multicolumn{3}{c}{4.7\%} & \multicolumn{3}{c}{13.1\%} & \multicolumn{3}{c}{13.7\%} & \multicolumn{3}{c}{1.9\%} & \multicolumn{3}{c}{0.6\%} & \multicolumn{3}{c}{12.2\%} & \multicolumn{3}{c}{40.9\%} & \multicolumn{3}{c}{14.8\%} & \multicolumn{3}{c}{30.8\%} \\ 

\specialrule{\lightrulewidth}{0pt}{0pt}

\textit{AlwaysOne} & 23\% & $\pm$ & 3\% & 14\% & $\pm$ & 3\% & 41\% & $\pm$ & 2\% & 12\% & $\pm$ & 2\% & 5\% & $\pm$ & 2\% & 11\% & $\pm$ & 4\% & 12\% & $\pm$ & 4\% & 2\% & $\pm$ & 1\% & & -- & & & -- & & & -- & & & -- & & & -- &          \\

\rowcolor{lightgray}
\textit{AlwaysZero} & 41\% & $\pm$ & 2\% & 46\% & $\pm$ & 1\% & 23\% & $\pm$ & 3\% & 46\% & $\pm$ & 1\% & 49\% & $\pm$ & 1\% & 47\% & $\pm$ & 1\% & 46\% & $\pm$ & 1\% & 49\% & $\pm$ & 1\% & & -- & & & -- & & & -- & & & -- & & & -- &          \\

\textit{UniformRandom} & 48\% & $\pm$ & 1\% & 43\% & $\pm$ & 2\% & 48\% & $\pm$ & 1\% & 42\% & $\pm$ & 1\% & 37\% & $\pm$ & 2\% & 42\% & $\pm$ & 3\% & 42\% & $\pm$ & 3\% & 35\% & $\pm$ & 1\% & 1.4\% & $\pm$ & 0.8\% & 16\% & $\pm$ & 3\% & 36\% & $\pm$ & 16\% & 18\% & $\pm$ & 4\% & 26\% & $\pm$ & 7\% \\

\rowcolor{lightgray}
\textit{InformedRandom} & 50\% & $\pm$ & 1\% & 50\% & $\pm$ & 1\% & 50\% & $\pm$ & 1\% & 50\% & $\pm$ & 1\% & 50\% & $\pm$ & 1\% & 50\% & $\pm$ & 1\% & 50\% & $\pm$ & 1\% & 50\% & $\pm$ & 1\% & 1\% & $\pm$ & 1.4\% & 14\% & $\pm$ & 3\% & 46\% & $\pm$ & 10\% & 16\% & $\pm$ & 4\% & 32\% & $\pm$ & 5\% \\

\specialrule{\lightrulewidth}{0pt}{0pt}
\rowcolor{white}
AudioOnly & 52\% & $\pm$ &1\% & 51\% & $\pm$ & 2\% & 53\% & $\pm$ & 3\% & 52\% & $\pm$ & 2\% & 50\% & $\pm$ & 1\% & 51\% & $\pm$ & 1\% & 51\% & $\pm$ & 2\% & 50\% & $\pm$ & 1\% & 0\% & $\pm$ & 0\% & 7\% & $\pm$ & 3\% & \highlight{53\%} & \highlight{$\pm$} & \highlight{4\%} & 15\% & $\pm$ & 4\% & \highlight{41\%} & \highlight{$\pm$} & \highlight{3\%} \\

\rowcolor{lightgray}
TextWithTiming & \highlight{59\%} & \highlight{$\pm$} & \highlight{3\%} & 50\% & $\pm$ & 2\% & \highlight{60\%} & \highlight{$\pm$} & \highlight{4\%} & \highlight{59\%} & \highlight{$\pm$} & \highlight{5\%} & \highlight{64\%} & \highlight{$\pm$} & \highlight{9\%} & \highlight{67\%} & \highlight{$\pm$} & \highlight{5\%} & \highlight{65\%} & \highlight{$\pm$} & \highlight{4\%} & 56\% & $\pm$ & 8\% & 0\% & $\pm$ & 0\% & 14\% & $\pm$ & 4\% & 47\% & $\pm$ & 3\% & 21\% & $\pm$ & 4\% & \highlight{41\%} & \highlight{$\pm$} & \highlight{4\%} \\

\rowcolor{white}

TextNoTiming        & \highlight{59\%} & \highlight{$\pm$} & \highlight{3\%} & 50\% & $\pm$ & 3\% & \highlight{58\%} & \highlight{$\pm$} & \highlight{3\%} & \highlight{57\%} & \highlight{$\pm$} & \highlight{3\%} & \highlight{63\%} & \highlight{$\pm$} & \highlight{7\%} & \highlight{67\%} & \highlight{$\pm$} & \highlight{5\%} & \highlight{65\%} & \highlight{$\pm$} & \highlight{5\%} & 59\% & $\pm$ & 10\%  & 0\% & $\pm$ & 0\%     & 14\% & $\pm$ & 6\% & 47\% & $\pm$ & 3\% & 20\% & $\pm$ & 5\% & \highlight{39\%} & \highlight{$\pm$} & \highlight{4\%} \\

\rowcolor{lightgray}
BothModalities & \highlight{59\%} & \highlight{$\pm$} & \highlight{3\%} & 50\% & $\pm$ & 2\% & \highlight{58\%} & \highlight{$\pm$} & \highlight{3\%} & \highlight{58\%} & \highlight{$\pm$} & \highlight{4\%} & \highlight{63\%} & \highlight{$\pm$} & \highlight{8\%} & \highlight{65\%} & \highlight{$\pm$} & \highlight{4\%} & \highlight{64\%} & \highlight{$\pm$} & \highlight{5\%} & 57\% & $\pm$ & 9\% & 0\% & $\pm$ & 0\% & 14\% & $\pm$ & 5\% & 47\% & $\pm$ & 3\% & 20\% & $\pm$ & 4\% & \highlight{40\%} & \highlight{$\pm$} & \highlight{3\%} \\
\bottomrule
\end{tabular}}
\caption{Gesture-property prediction scores for all baselines, and our trained predictors using text, audio, or both modalities. Baselines are \textit{italicised}; \highlight{\fbseries{bold, coloured} numbers} indicate that the given label is found to be predictable as defined in Sec.\ \ref{ssec:baselines}.}
\label{tab:speech_modalities}
\end{table*}
\renewcommand{\arraystretch}{1}
\normalsize
\begin{table*}[t]
\centering
%\rowcolors{3}{}{lightgray} % Coloured rows
% Fit the whole page
\resizebox{\textwidth}{!}{
% Taller lines
\renewcommand{\arraystretch}{1.5}
% Magic for aligning on plus-minus symbol, see property_prediction.text
\begin{tabular}{@{\kern\tabcolsep}r *{39}{r@{\hspace{\tabcolsep}}c@{\hspace{\tabcolsep}}l}@{\kern\tabcolsep}}

\specialrule{\heavyrulewidth}{0pt}{0pt}&
\multicolumn{12}{c}{\textbf{gesture category} [Macro $\text{F}_1$]} & \multicolumn{12}{c}{\textbf{gesture semantics} [Macro $\text{F}_1$]} & \multicolumn{15}{c}{\textbf{gesture phase} [$\text{F}_1$]} \\

\cmidrule(lr){2-13}
\cmidrule(lr){14-25}
\cmidrule(lr){26-40}

label & \multicolumn{3}{c}{deictic} & \multicolumn{3}{c}{beat} & \multicolumn{3}{c}{iconic} & \multicolumn{3}{c}{discourse} & \multicolumn{3}{c}{amount} & \multicolumn{3}{c}{shape} & \multicolumn{3}{c}{direction} & \multicolumn{3}{c}{size} & \multicolumn{3}{c}{pre-hold} & \multicolumn{3}{c}{post-hold} & \multicolumn{3}{c}{stroke} & \multicolumn{3}{c}{retraction} & \multicolumn{3}{c}{preparation} \\

\specialrule{\lightrulewidth}{0pt}{0pt}
BetweenSpeaker & 58\% & $\pm$ & 3\% & 49\% & $\pm$ & 2\% & 56\% & $\pm$ & 5\% & 56\% & $\pm$ & 4\% & 61\% & $\pm$ & 5\% & 67\% & $\pm$ & 7\% & 65\% & $\pm$ & 6\% & 59\% & $\pm$ & 11\% & 0\% & $\pm$ & 0\% & 12\% & $\pm$ & 8\% & 42\% & $\pm$ & 7\% & 20\% & $\pm$ & 6\% & 38\% & $\pm$ & 6\% \\

\rowcolor{lightgray}
WithinSpeaker & 59\% & $\pm$ & 3\% & 50\% & $\pm$ & 2\% & 58\% & $\pm$ & 3\% & 58\% & $\pm$ & 4\% & 63\% & $\pm$ & 8\% & 65\% & $\pm$ & 4\% & 64\% & $\pm$ & 5\% & 57\% & $\pm$ & 9\% & 0\% & $\pm$ & 0\% & 14\% & $\pm$ & 5\% & 47\% & $\pm$ & 3\% & 20\% & $\pm$ & 4\% & 40\% & $\pm$ & 3\% \\

\rowcolor{white}
$\text{WithinSpeaker}_{\text{ID}}$ & 60\% & $\pm$ & 2\% & 53\% & $\pm$ & 3\% & 62\% & $\pm$ & 4\% & 61\% & $\pm$ & 3\% & 64\% & $\pm$ & 9\% & 67\% & $\pm$ & 7\% & 64\% & $\pm$ & 5\% & 54\% & $\pm$ & 7\% & 0\% & $\pm$ & 0\% & 12\% & $\pm$ & 5\% & 54\% & $\pm$ & 5\% & 22\% & $\pm$ & 7\% & 45\% & $\pm$ & 3\% \\

\specialrule{\heavyrulewidth}{0pt}{0pt}
\end{tabular}}
\caption{A comparison of prediction results for different cross-validation strategies. For detailed information see \cref{sec:generalization_study}.}
\label{tab:generalisation}
\end{table*}

This experiment analysed the importance of the two input speech modalities -- text and audio (prosodic features) -- for predicting the gesture properties under study. We consider four different versions of our model: trained using only speech prosodic features (\textit{AudioOnly}); trained using only text features, with explicit timing information for each word either included (\textit{TextWithTiming}) or omitted (\textit{TextNoTiming}); and trained on both speech audio and text features, including word timing (\textit{BothModalities}).

As can be seen from the results in Table \ref{tab:speech_modalities}, the text features were informative 
%contains all the helpful information
and predict gesture category and gesture semantics content better than chance.
Audio features, in contrast, did not improve on the best bottom-line predictions for these gesture properties, and so did not allow better-than-chance prediction on their own. This provides evidence that audio features alone are insufficient for predicting the semantics of iconic gestures. It also indicates that even the category of the gesture cannot be predicted from audio features alone.
Moreover, the combination of audio and text did not in general perform better than text on its own.
Text appears to be a necessary input for both gesture category and semantics. This has strong implications for the need to include text in machine learning gesture models in general.
This finding can be explained by semantic information, which gesture category strongly depends on, being challenging to obtain directly from the audio. Whether or not an iconic gesture is appropriate will generally depend on the semantic information in the speech.

A different story emerges for gesture phase prediction, where audio was more helpful than text (even with word timing information). %, as can be seen in Table \ref{tab:SpModPhase}.
In particular, gesture stroke could be predicted noticeably better than informed random sampling using audio.
%Audio showed the best performance with the least variance, but text was only two percent lower.
%The utility of text may relate to the relationship between the verbal message and prosodic prominence -- prosody is predictable from text, as TTS systems show (although their input is graphemic symbols rather than semantic word vectors). Our default text features (used by \emph{Text\-With\-Timing} and \emph{Both\-Modalities}) also include word timing information relative to the current frame, but this does not seem essential since a system without these numbers (\emph{Text\-No\-Timing}) performed equally well.
%Our representation of each word also contains information about the time difference from the current frame, and this information could potentially be utilised for gesture phase prediction, even though our results show that even without the timing information text was equally useful.
This confirms previous findings that gesture stroke can be predicted from speech audio \cite{yunus2020sequence}. %, but sheds new light on which modality is needed for such prediction: while previously speech audio was mainly used, time-aligned speech text could also be helpful on its own.

\subsection{Generalising within and across speakers}
\label{sec:generalization_study}

Next, we evaluated gesture property prediction performance when generalising to novel speakers, versus the performance on speakers present in the training data.

Table \ref{tab:generalisation} compares prediction results
of a hold-one-speaker-out cross-validation strategy
%when the test set contains only a novel speaker whose data was not used for training
(\emph{BetweenSpeaker}) to our default cross-validation strategy where speakers are present in both training and test sets (\emph{WithinSpeaker}), and when speakers are present in both sets and the model has access to speaker IDs encoded as one-hot vectors (\emph{$\text{WithinSpeaker}_{\text{ID}}$}). 

We observe minor performance changes when we evaluate on previously unseen speakers, but the changes are not substantial and performance remains better than chance. %The tables indicate the same tendency for the other gesture properties.
As gesture behaviour is highly idiosyncratic across individuals, one might a priori have expected a large drop in prediction performance.
%it seems reasonable to expect a priori that there might be a large drop in between speaker generalisation.
It is reassuring that the actual change is quite modest, suggesting that gesture-property predictors generalise relatively well to new speakers.

Conversely, we observe no notable difference between \emph{WithinSpeaker} and \emph{$\text{WithinSpeaker}_{\text{ID}}$}, indicating that predictions did not benefit from knowing the speaker label. One reason could be that each speaker only has 10 min of data, which may not allow learning personalised correlations.
As above, another related contributing factor could be that
%Another hypothetical reason could be that we do not learn much speaker-specific information because
gesture behaviour is very stochastic overall.

We have also investigated the generalisation gap between training and validation performance. %As can be seen in our results on FigShare, \href{https://doi.org/10.6084/m9.figshare.15134796}{doi.org/10.6084/m9.figshare.15134796}, 
For brevity, the exact numbers are not reported, but the model overfits by double digits for gesture semantics and category, but not for gesture phase.
It may be that gesture-phase prediction is ambiguous and difficult even on training data with the features we used.
%our low-dimensional audio features.
%It seems to indicate that predicting the gesture phase from the speech is so difficult that the model does not perform well even on the training data.

\subsection{Some prediction examples}

\begin{figure}
    \centering
    \includegraphics[width=0.4\textwidth]{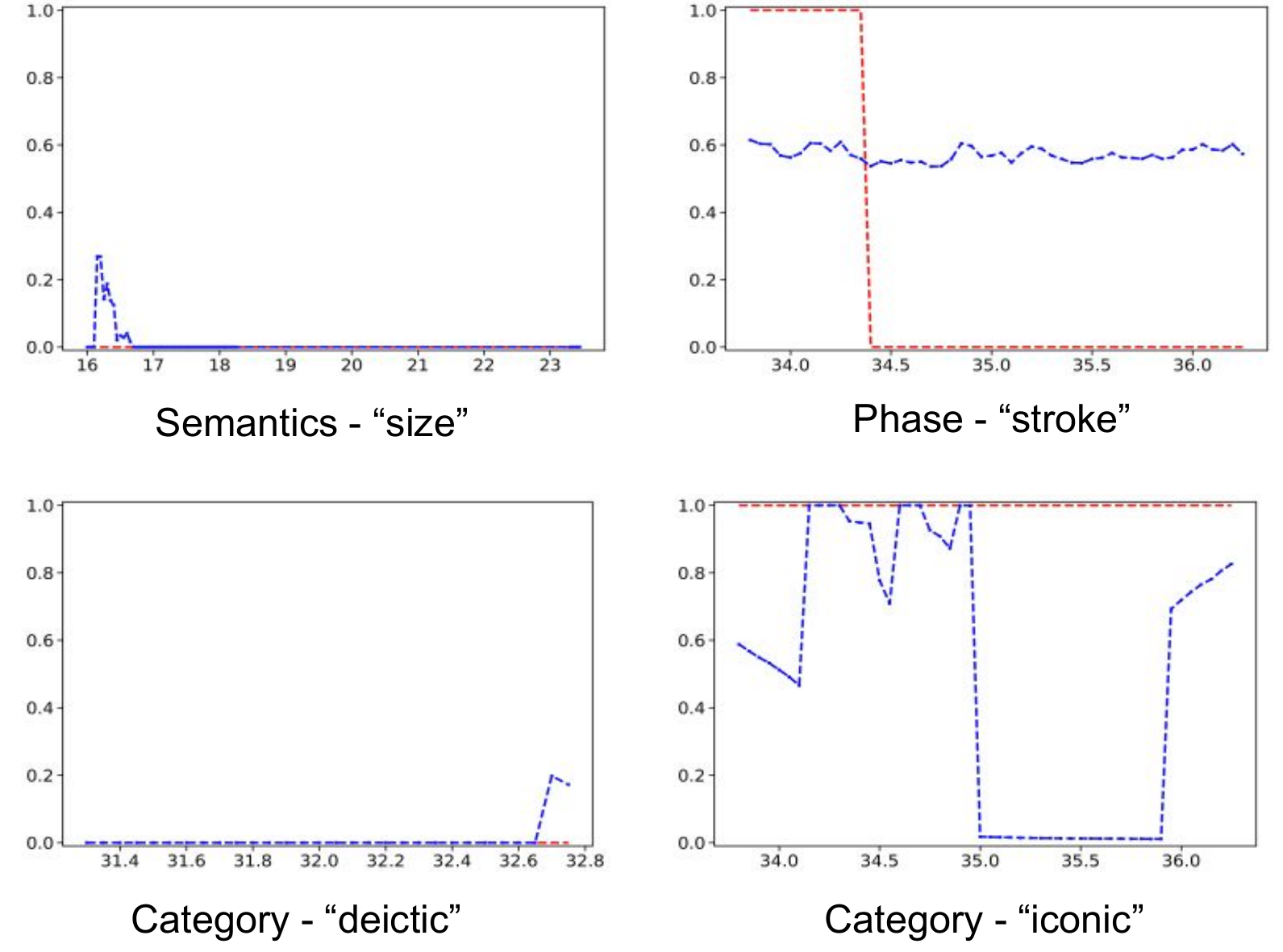}
    \caption{Example output sequences from gesture label prediction. The $x$-axes are time (in seconds) and the $y$-axes label probability. True labels are red, predictions blue.}
    \label{fig:pred_results}
    \Description{Four line plots showing, for 4 different labels, the predicted probabilities and the ground-truth for particular input windows lasting several seconds. 
    
    For "semantics - size", the ground truth is 0 throughout the 5 second long input window, and the example prediction  is around 0 percent with a small peak going up to 20 percent then down to 0 again in the first second of the input.
    
    The ground truth and the example prediction for "category - deictic" is similar, except the prediction has its peak at the end of the window. 
    
    The ground truth for "category - iconic" is always 1 over the 3 second long window, but the prediction is more erratic. In the first half of the input, the prediction starts from 60 percent, goes down to 50 percent, then quickly rises up and starts shifting between 80 and 100 percent. In the second half, it suddenly drops to zero and stays there for a second, then it instantly goes up to 70 percent and rises up to 80 percent in the last half second.
    
    For "phase - stroke", the ground truth is 1 in the first second and then zero in the remaining 2 seconds, but the prediction, hovers between 50 to 60 percent throughout the entire window.}
\end{figure}

Figure \ref{fig:pred_results} shows several example sequences of per-frame predicted probabilities for the presence of various gesture property labels on a held-out speech sequence. We can see both good and bad performance.
%Interestingly, predictions for the stroke seem to be slightly shifted in time.
%In general, predicted sequences are somewhat jagged, suggesting that future predictors could benefit from temporal smoothing or a more explicit model of labels over time.
We note that the predictions for gesture semantics and category are surprisingly confident, given that cross-entropy tends to promote cautious models that favour predicting numbers close to the a-priori class probabilities in the absence of compelling evidence to the contrary.
This behaviour might be due to overfitting.
%Several other example plots are attached as supplementary material.

\subsection{Do all speakers have similar predictability?}

\begin{figure}
\centering
  \includegraphics[width=0.9\columnwidth]{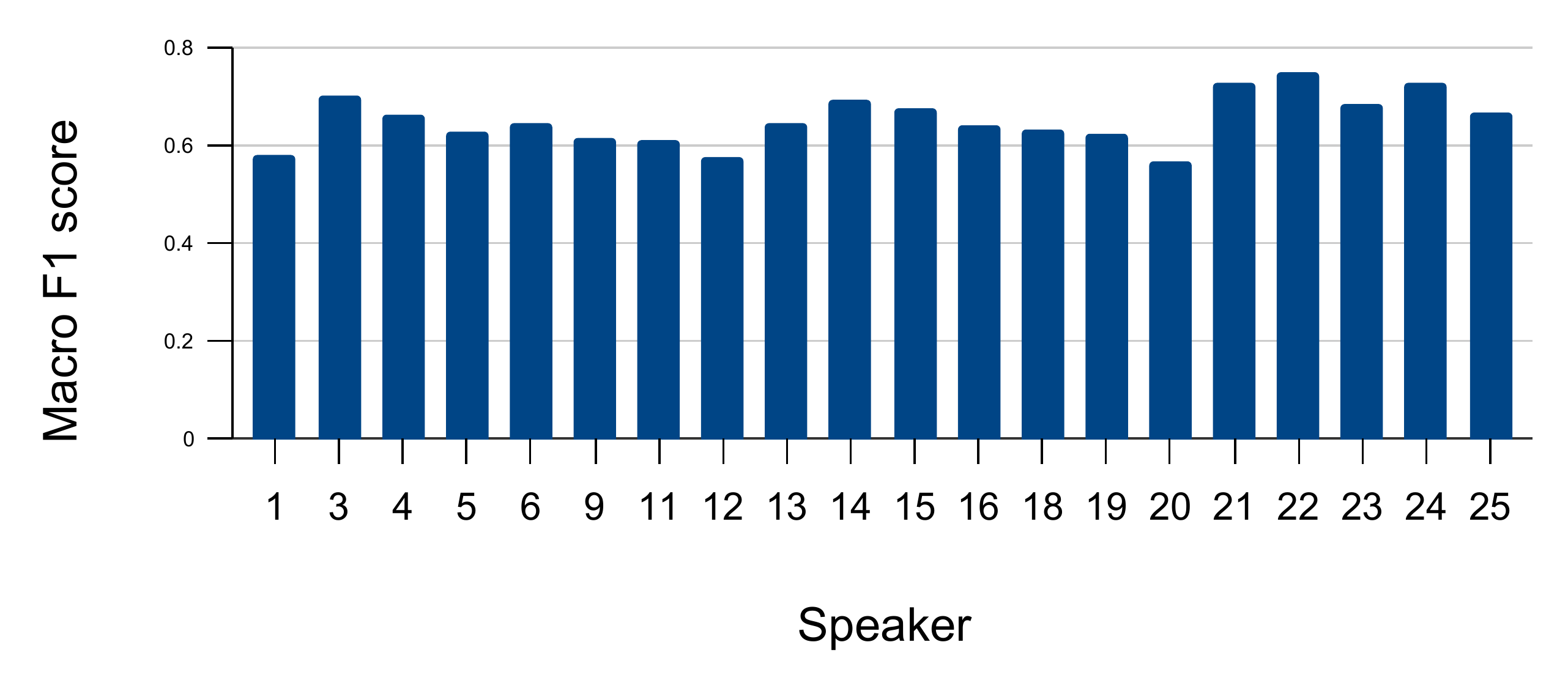}
  \caption{Macro $\text{F}_1$ score for predicting the ``direction'' label across the 22 speakers considered. Note the high variation.}
  \label{fig:speaker_variability}
  \Description{A bar plot showing 22 Macro F1 scores. The scores wildly vary between roughly 58 percent and roughly 76 percent. }
\end{figure}

%Here we talk about variability between speakers.

There is a large difference in prediction performance for different speakers, as seen in the high standard deviation in most of the tables above. We show one example of the performance variation between speakers in Figure \ref{fig:speaker_variability}. We see that how well we predict the gesture semantic label ``direction'' changes substantially from speaker to speaker, indicating that not all speakers are equally predictable, although predictions are better than chance in all these cases.

\subsection{On the effect of hyperparameters}

\begin{figure}
\centering
  \includegraphics[width=0.45\textwidth]{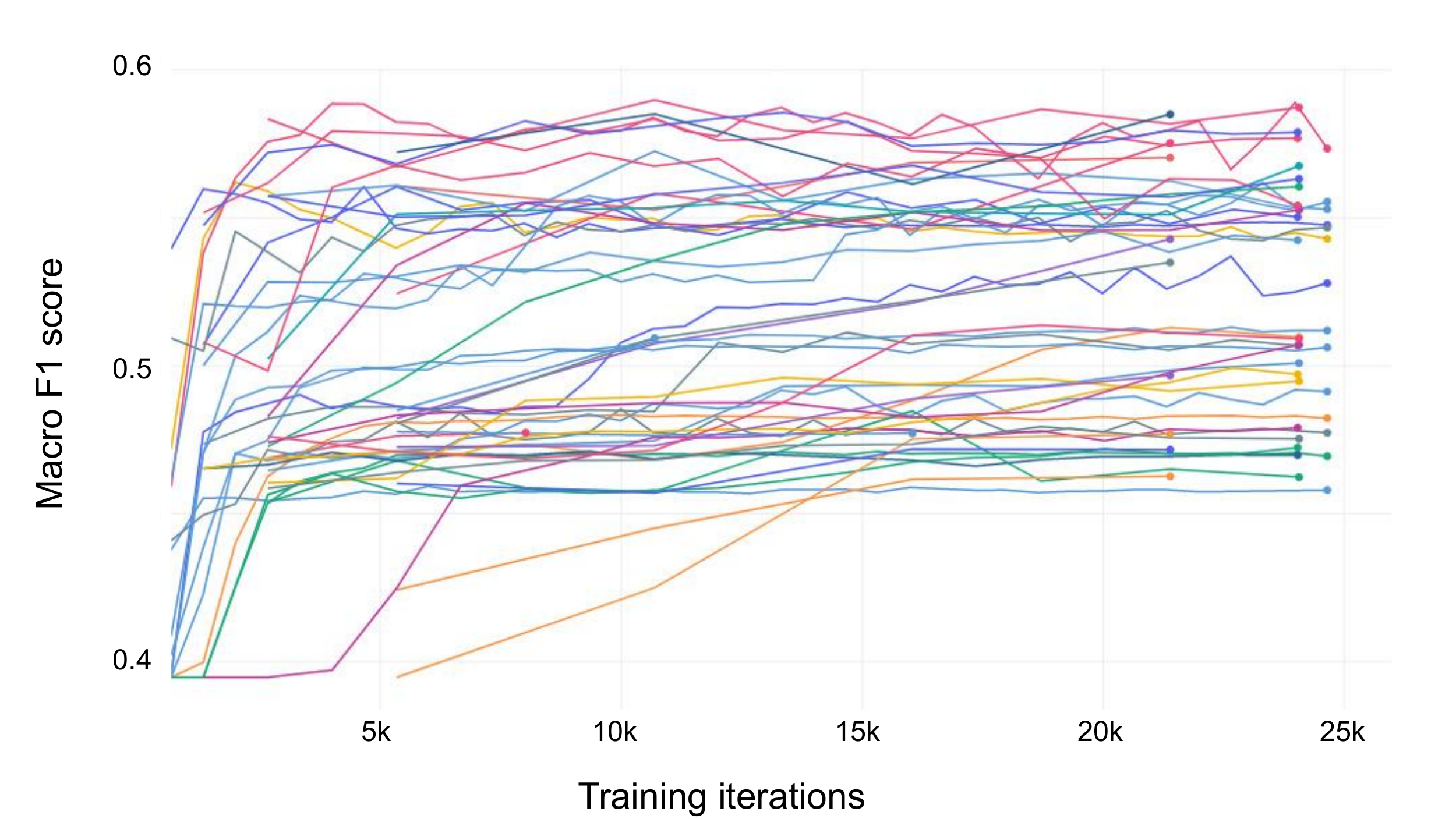}
  \caption{Training evolution of gesture category prediction average Macro $\text{F}_1$ score for 50 different hyperparameter settings. The $x$-axis shows the number of update steps and the $y$-axis the average Macro $\text{F}_1$ score. The total number of training steps was fixed to 25k.}
  \label{fig:hparams_variability}
  \Description{A line plot with 50 lines that show the evolution of the macro F1 during training. It shows that the training is relatively stable for individual hyperparameter settings (i.e., the macro F1 rises slowly until it converges), but the performance heavily depends on the hyperparameter setting: the final performances are roughly evenly distributed between 46 and 58 percent. The difference between the hyperparameter settings tends to already manifest at 5 thousand out of 25 thousand training iterations.}
\end{figure}

Aside from the large variation between different speakers, we also observe substantial performance variation depending on model hyperparameters. Figure \ref{fig:hparams_variability} shows the Macro $\text{F}_1$ score for predicting gesture category for 50 different hyperparameter runs, as described in Section~\ref{ssec:model}.
We can see that results vary greatly depending on hyperparameters.
The performance variation attributable to the hyperparameters is much greater than the difference between many conditions in our experiments, indicating that the model is sensitive to the hyperparameter settings.
We recommend future work to also perform hyperparameter search to obtain reliable performance.

\section{Conclusions}
We have studied the extent to which 13 different gesture-property labels -- mainly ones of relevance to communicative gestures -- can be predicted from speech.
Numerous experiments on a direction-giving dataset
%In this paper, we conducted numerous gesture-property prediction experiments on a direction-giving dataset. Our results
show that the gesture properties we considered, such as gesture semantics and gesture phases, can be predicted from speech with Macro $\text{F}_1$ scores better than chance. Predicting gesture properties for speakers outside the training data was only slightly more challenging, suggesting that gesture-property prediction may generalise well.

%Human gesticulation is highly stochastic, so the maximum achievable score is likely significantly below 100\%.

Another central finding is that, for predicting gesture properties such as gesture category and gesture semantics, all that must be known is the text transcript, while for others, specifically phase, prosodic audio features are much more suitable.
%should be used in place of text. 

%Although the advantage over the chance baselines is numerically small, this
Our $\approx$10\% advantage over the chance baselines
should be viewed in the context that human gestures are highly stochastic, and that state-of-the-art data-driven gesture synthesis does not compare favourably to random gesticulation \cite{kucherenko2020genea}. Leveraging our gesture-property predictions to achieve semantically appropriate gestures even a fraction of the time could thus add important communicative value,
%Since state-of-the-art data-driven gesture synthesis does not produce meaningful gesticulation \cite{kucherenko2020genea}, being able to replace current, ``low meaning'' gestures with semantically appropriate ones even a fraction of the time could add important communicative value.
and the prediction models we identify are well-suited for integration into modern data-driven gesture generation systems.
%to synthesise gestures with rich semantics.
%Future, larger datasets and additional prediction approaches may improve scores and potentially also render other gesture properties predictable, but cannot take away from our findings that certain properties are possible to predict.

\subsection{Future work}
The present study opens up several directions for future research:

%First, there are many alternative design choices left to explore, e.g., 
First, the study could be expanded, e.g., to perform more extensive architecture search and evaluate on a metric that (like human perception \cite{nirme2019motion}) is less sensitive to small timing shifts than the measures in this article, which compare each frame individually.

Second, it would be interesting to perform a similar study on other datasets, e.g., in different languages or from situations other than direction giving. Also, while the SaGA dataset we used is the largest one we know that has been annotated at this high level of detail, larger datasets are also of interest as they become available.

Last, an important future goal is integrating gesture-property prediction into modern gesture-generation models, to enable more appropriate and meaningful gesture synthesis, as for example suggested in~\cite{kucherenko2021speech2properties2gestures}.

%Second, it would be interesting to apply the same study to data in other languages, to investigate to what extent the findings are tied to German. Moreover, it is relevant to study settings different from the direction-giving scenario investigated here.

%Third, it would also be valuable to evaluate the generated gestures using a metric that is less sensitive to small timing shifts.
%, where the present metric might be unnecessarily strict.
%Human perception of co-speech gesture naturalness is not very sensitive to small gesture-timing deviations \cite{nirme2019motion}, but the frame-wise metrics in this article require perfect alignment for maximum score.

%The present metric might be overly strict concerning timing. Generated examples that are deemed as wrong by this metric might actually be natural, as we know that humans are not very sensitive to small gesture timing deviations in their perception of co-speech gesture naturalness \cite{nirme2019motion}.

%\section*{Acknowledgement}
\begin{acks}
The authors are grateful to Stefan Kopp for providing the SaGA dataset and fruitful discussions about it, and to Olga Abramov for advising on the dataset gesture-property processing.
This work was partially supported by the Swedish Foundation for Strategic Research Grant No.\ RIT15-0107 and by the Wallenberg AI, Autonomous Systems and Software Program (WASP) funded by the Knut and Alice Wallenberg Foundation.
\end{acks}

\bibliographystyle{ACM-Reference-Format}
\bibliography{sample-bibliography,sample}

\end{document}